\documentclass[twocolumn,showpacs,preprintnumbers,amsmath,amssymb]{revtex4}
\usepackage{epsfig}

\usepackage{graphicx}
\usepackage{dcolumn}
\usepackage{bm}

\usepackage{amsthm}

\begin{document}


\title{Comment on ``Arbitrated quantum-signature scheme"}

\author{Marcos Curty$^{1}$ and Norbert L\"{u}tkenhaus$^{2,3}$}
\affiliation{ $^1$ ETSI Telecomunicaci\'on, 
University of Vigo, Campus Universitario, 36310 Vigo, Spain \\
$^2$ Institute for Quantum Computing, University of Waterloo, 200
University Avenue West, Waterloo, Ontario N2L 3G1, Canada\\ $^3$
Quantum Information Theory Group, Institut f\"ur Theoretische
Physik I, and Max-Planck Research Group, Institute of Optics,
Information and Photonics, Universit\"{a}t Erlangen-N\"{u}rnberg,
Staudtstra{\ss}e 7/B2, 91058 Erlangen, Germany}


\begin{abstract}
We investigate the quantum signature scheme proposed by Zeng and
Keitel [Phys. Rev. A \textbf{65}, 042312 (2002)]. It uses
Greenberger-Horne-Zeilinger (GHZ) states and the availability of
a trusted arbitrator. However, in our opinion the protocol is not clearly
operationally defined and several steps are ambiguous. Moreover, we 
argue that 
the security statements claimed by the authors are incorrect.
\end{abstract}


\maketitle

Digital signature schemes provide message authentication which
enables third parties to settle disputes about the authenticity
of messages. In Ref.~\cite{ZENG_2002} Zeng and Keitel proposed a
quantum signature scheme that requires the availability of a
trusted arbitrator as part of the signature initialization and
verification algorithms. In our opinion, the protocol is not well
operationally defined, its presentation is misleading, and
several steps are ambiguous. Moreover, we believe that the
security statements claimed by the authors are incorrect. We
first list the main points of our criticism and then provide more
details.

The scheme proposed in Ref.~\cite{ZENG_2002} has as its goal to sign a quantum
state $|P\rangle$. From the paper, however, it is not clear whether the 
sender (Alice), the
receiver (Bob), or the arbitrator, need to know the identity of
the quantum state $|P\rangle$ to be signed, or whether they have
access to a restricted number of copies of an unknown state
$|P\rangle$. One of the main motivations for the work presented in
Ref.~\cite{ZENG_2002} is that ``classical signature schemes are
difficult to assign to messages in qubit format''. Then, one
might be tempted to assume that none of the parties involved in
the communication has a classical description of the state
$|P\rangle$. However, it is well known that signing unknown
quantum messages is not possible \cite{BARNUM}. One can then
consider that all the parties know the state $|P\rangle$. This
assumption renders the quantum signature scheme proposed in
Ref.~\cite{ZENG_2002} to one intended to sign classical data
using quantum resources. However, in this scenario it is unclear
what are the real advantages of this protocol, if any, with
respect to unconditionally secure classical signature schemes
(see, e.g., Ref.~\cite{CHAUM_1991} and references therein).
Finally, one can assume the natural scenario where Bob (or even
the arbitrator) do not know the state $|P\rangle$. However, as we
will show below, the signature scheme proposed 
in Ref.~\cite{ZENG_2002} is insecure in this last case \cite{note1}.

In several crucial points of the protocol a step of
state comparison is required . Especially, if Bob does not know the state
$|P\rangle$ to be signed, he would have to compare two unknown states. 
The authors of Ref.~\cite{ZENG_2002}
did not clarify
in their manuscript how to perform these quantum state comparison
steps, and they treat them as deterministic and
error-free processes. However, it is evident from the no-cloning theorem 
\cite{noclon}
to be impossible to do
universal quantum state comparison in a deterministic way and without 
disturbing the original states. For a quantitative analysis of this scenario 
see Ref.~\cite{BARNETT_2002}, where the optimal 
comparison test and its success 
probability have been recently obtained.

Let us now discuss our criticism in more detail. We start with a
brief description of the protocol. The scheme includes three
phases \cite{ZENG_2002}: an initial phase, a signing phase, and a
verification phase. In the first one, Alice, Bob, and the
arbitrator distribute two secret keys, $K_a$ (Alice-arbitrator)
and $K_b$ (Bob-arbitrator). These two secret keys might consist
of quantum states or of classical data. Next, they create and
distribute GHZ states. The distribution of GHZ states has to be
repeated for every single communication: the ``algorithm relies
crucially on the entanglement of the three involved
communicators''. In this Comment we will consider that this
initial phase can be completed in a safe manner, although the 
authors of Ref.~\cite{ZENG_2002} do not present any specific protocol 
to verify the correct execution of entanglement distribution.

The signing phase can be used to sign {\it pure} $n$-qubit messages of
the form
$|P\rangle=\bigotimes_{i=1}^n(\alpha_i|0\rangle+\beta_i|1\rangle)$.
The signature of $|P\rangle$, denoted as $|S\rangle$, is defined as a
quantum encryption of some classical data $\cal{M}$$_a$ and a
quantum state $|R\rangle$.
In order to encrypt
this information Ref.~\cite{ZENG_2002} proposes to use the ``approach known
as `quantum state operation'". It remains unclear what the authors 
mean with ``quantum state operation", but a quantum
one-time-pad scheme might be used for this purpose \cite{BOYKIN}.
More important, in this step it is not clearly defined how the
crucial quantum state $|R\rangle$ is generated by Alice. First,
it seems that Alice uses $K_a$ to select a set of ``measurement
operators'' $\cal{M}$$_{K_a}$. If $K_a$ denotes a quantum state
$|K_a\rangle$, then $\cal{M}$$_{K_a}$ must contain $|K_a\rangle$
as an eigenvector. Note, however, that in this scenario it
remains unclear how Alice can obtain $\cal{M}$$_{K_a}$ from
$|K_a\rangle$ if she does not have a classical description of the
quantum state $|K_a\rangle$. If $K_a$ is a classical key then
$\cal{M}$$_{K_a}$ can represent any measurement operator within a
given set indexed by $K_a$. Next, the sentence ``Alice is
required to measure the information string of qubits $|P\rangle$
using $\cal{M}$$_{K_a}$ and obtains $|R\rangle$'' seems to
indicate that $|R\rangle$ arises from a measurement on
$|P\rangle$, {\it i.e.}, $\cal{M}$$_{K_a}$ is an observable.
Note, however, that in this case the protocol can only work
probabilistically. After receiving an advance copy of this
manuscript, the authors of Ref.~\cite{ZENG_2002} emphasized that
$\cal{M}$$_{K_a}$ denotes a unitary transformation
\cite{ZENG_PRIVATE}. Therefore, from now on we will consider that
$|R\rangle=\cal{M}$$_{K_a}|P\rangle$ with $\cal{M}$$_{K_a}$
unitary.


The verification algorithm requires the arbitrator to obtain a
parameter $\gamma$ arising from a forgery test. In
order to do that, he needs to generate two quantum states,
$|R\rangle$ and $|R'\rangle$, that need to be compared (Step $2$
in the verification phase). If $|R\rangle$ and $|R'\rangle$ are
different, then $\gamma=0$ and $|P\rangle$ has to be rejected.
Otherwise $\gamma=1$ and Bob needs to perform a second
verification test. Here, again, it is not clearly stated how does
the arbitrator obtain these two states $|R\rangle$ and
$|R'\rangle$ from $\cal{M}$$_b$, $|S\rangle$, and $|P\rangle$ sent
by Bob. More important, as pointed out above, the authors of 
Ref.~\cite{ZENG_2002} do
not explain how the quantum state comparison test between
$|R\rangle$ and $|R'\rangle$ is performed. 

Once the first forgery test introduced above concludes, the
arbitrator needs to obtain a parameter $\cal{M}$$_t$. The
procedure to generate $\cal{M}$$_t$ is a bit misleading . 
Ref.~\cite{ZENG_2002} claims that ``the arbitrator
measures or evaluates the states of the particles in his string
of GHZ states''. Again, here it is not clear the meaning of
``evaluates''. Once $\cal{M}$$_t$ is obtained, whatever the
process involved, the arbitrator prepares a quantum state
$y_{tb}$ containing part of the information obtained in the
previous steps of the protocol and he sends it to Bob.

Depending on the contents of $y_{tb}$, Bob needs
to decide whether the message originates from Alice or not.
This constitutes the last step of the verification phase. Now Bob
has to compare the quantum state $|P\rangle$ with a state $|P'\rangle$. ``If
$|P'\rangle=|P\rangle$, the signature is completely correct and
Bob accepts $|P\rangle$, otherwise, he rejects it''. Again, at
this crucial point we find the problem of how to obtain
the quantum states $|P\rangle$ and $|P'\rangle$ from $y_{tb}$,
and how to realize the quantum state comparison test.


Next, we show that the protocol presented in Ref.~\cite{ZENG_2002}
cannot lead to a secure signature scheme if Bob and the arbitrator 
do not know the state
$|P\rangle$. To simplify our notation, we shall mainly consider
one-qubit messages, {\it i.e.},
$|P\rangle=\alpha{}|0\rangle+\beta{}|1\rangle$.

To obtain the parameter $\cal{M}$$_a$, Alice performs a Bell
measurement on a copy of the state $|P\rangle$ and her particle of
the GHZ state. 
Let us assume, for instance, that $\cal{M}$$_a$ corresponds to
the state $|\Psi_{12}^-\rangle_a$ (see Eq.~(8) in
Ref.~\cite{ZENG_2002}), which will always occur with probability
$1/4$, and that $|R\rangle$ has been obtained as
$|R\rangle=\cal{M}$$_{K_a}|P\rangle$, with $\cal{M}$$_{K_a}$
denoting a unitary transformation. The correlations of the GHZ
state impose, in this case, that the state shared by Bob and the
arbitrator is $|\varphi\rangle=\alpha|00\rangle-\beta|11\rangle$.

The verification phase begins once Bob receives $|P\rangle$ and
$|S\rangle$ from Alice. Here Bob measures his particle of
$|\varphi\rangle$ in the $x$ direction. The result is
recorded in the parameter $\cal{M}$$_b$. The state
$|\varphi\rangle=\alpha|00\rangle-\beta|11\rangle$ can be
written as
$|\varphi\rangle=1/\sqrt{2}(|+x\rangle\sigma_Z|P\rangle+|-x\rangle|P\rangle)$,
where $|\pm{}x\rangle=1/\sqrt{2}(|0\rangle\pm{}|1\rangle)$, and
$\sigma_Z$ is the Pauli matrix ($\sigma_Z|0\rangle=|0\rangle$ and
$\sigma_Z|1\rangle=-|1\rangle$). Both possible results,
$\{|\pm{}x\rangle\}$, have equal a priori probability $1/2$. Let
us consider, for instance, that $\cal{M}$$_b=|+x\rangle$. The
state of the arbitrator's particle is then reduced to
$\sigma_Z|P\rangle$.

Next, Bob sends $y_b=K_b(\cal{M}$$_b,|S\rangle,|P\rangle)$ to the
arbitrator. With this information the arbitrator performs his
forgery test. Now, in order to obtain $|R\rangle$ and
$|R'\rangle$, we consider two possible alternatives. On the one
hand, to evaluate if the message received by Bob is authentic, it
seems that $|R\rangle$ and $|R'\rangle$ should depend on
$|P\rangle$ and $|S\rangle$. That is, $|R\rangle$ originates from
$|P\rangle$ as $|R\rangle=\cal{M}$$_{K_a}|P\rangle$, and
$|R'\rangle$ from $|S\rangle$ (or vice versa). On the other hand,
the authors of Ref.~\cite{ZENG_2002}) 
assert that the decryption of $|S\rangle$ ``gives
rise to $|R'\rangle$ via the correlations of the GHZ state''. One
might then also think that $|R'\rangle$ (or $|R\rangle$) arises
from the GHZ particle of the arbitrator, and $|R\rangle$ (or
$|R'\rangle$) from $|S\rangle$ or $|P\rangle$. Note that by using
the correlations $\cal{M}$$_a$ (contained in $|S\rangle$), and
$\cal{M}$$_b$ the arbitrator can find that his particle is in the
state $\sigma_Z|P\rangle$. Then he could recover $|P\rangle$ 
applying $\sigma_Z$ ($\sigma_{Z}^2=I$).
More important, once the arbitrator obtains $|R\rangle$ and
$|R'\rangle$, whatever the process involved, he needs to compare
these two unknown quantum states to decide whether they are equal
or not. Unfortunately, 
it is known that it is impossible to
conclusively identify two pure unknown states as being identical
\cite{BARNETT_2002}. Nevertheless, one can perform a measurement
that examines whether the systems are not the same
\cite{BARNETT_2002}. Let $q$ denote the average success
probability of identifying two pure unknown states as different.
With this comparison procedure no valid messages will produce
$\gamma=0$, but we find that a forged message will be accepted
with probability $1-q$. For one-qubit messages we
have that $q=1/4$ \cite{BARNETT_2002}. Here we
consider that $|R\rangle$ and $|R'\rangle$ are selected at random
within the set of all pure states.  For $n$-qubit messages the
value of $q$ depends on $\cal{M}$$_{K_a}$. 
Ref.~\cite{ZENG_2002} seems to consider
$\cal{M}$$_{K_a}=\bigotimes_{i=1}^n\cal{M}$$^i_{K_a^i}$, with
$\cal{M}$$^i_{K_a^i}$ unitary for all $i$. 
$|P\rangle|S\rangle=\bigotimes_{i=1}^n|p_i\rangle|s_i\rangle$,
Now a potential adversary
could follow, for instance, an strategy that do not modify all
the $n$ qubits contained in $|P\rangle$, but only a small
fraction $m$ of them. This is sufficient to achieve a dramatic
decrease of the quantum fidelity \cite{FIDELITY} of the resulting
quantum state with respect to the original message $|P\rangle$.
In the worse-case-scenario ($m=1$) the arbitrator will accept a
forged message with probability $3/4$. One can improve the
ability of detecting forged messages by using a general unitary
transformation $\cal{M}$$_{K_a}$. Unfortunately, even for this
scenario the value of $q$ is relative low: $q=1/2\
(1-2^{-n})$ \cite{BARNETT_2002}. 
As a consequence, we find that a possible attacker
(which includes as well a potential dishonest Bob) could modify
Alice's messages such that the acceptance parameter $\gamma$
satisfies $\gamma=1$ with non negligible probability. Moreover,
note that so far we always assumed that $|R\rangle$ and
$|R'\rangle$ are pure states. A better security analysis 
against an
adversary that sends mixed states, would also be necessary here.

From now on, we shall presume that Bob is honest and we evaluate
his forgery test. For simplicity, we will consider
that the comparison process described above can be accomplished
without disturbing the original states.

After calculating $\gamma$, the arbitrator needs to obtain the
parameter $\cal{M}$$_t$. 
``Note
that $\cal{M}$$_t$ may be $|+x\rangle$ or $|-x\rangle$''.
It seems, therefore, that to obtain
$\cal{M}$$_t\in\{|\pm{}x\rangle\}$ the authors of Ref.~\cite{ZENG_2002} 
require that
the arbitrator measures his particle of the GHZ state in the $x$
direction. Furthermore, in Ref.~\cite{ZENG_2002} it is
specifically mentioned that ``the arbitrator may choose an
appropriate sequence of measurement operators to measure his GHZ
particle''. Once this measurement is performed, the arbitrator
sends Bob the state
$y_{tb}=K_b(\cal{M}$$_a,\cal{M}$$_b,\cal{M}$$_t,\gamma,|S\rangle)$.

Bob does not know Alice's secret key $K_a$. This means that from
$y_{tb}$ he cannot obtain the message $|P\rangle$ anymore. Note
that $|P\rangle$ cannot be calculated from $\cal{M}$$_a$,
$\cal{M}$$_b$, and $\cal{M}$$_t$ alone: the parameters
$\cal{M}$$_a$ and $\cal{M}$$_b$ are completely independent of
$|P\rangle$, whereas $\cal{M}$$_t=|\pm{}x\rangle$ only means that
$|P\rangle$ is not orthogonal to $|\mp{}x\rangle$ (assuming that
the arbitrator's particle was $\sigma_Z|P\rangle$). To avoid this
problem in the protocol, let us assume for the moment that
$y_{tb}$ also includes the message $|P\rangle$, or that Bob can
have access to a copy of the state $|P\rangle$ somehow.

Now the last step of the verification phase takes place. Here Bob
has to compare $|P\rangle$ with a state $|P'\rangle$. ``If
$|P'\rangle=|P\rangle$, the signature is completely correct and
Bob accepts $|P\rangle$, otherwise, he rejects it''. In order to
obtain $|P'\rangle$ Bob must use the parameters $\cal{M}$$_a$,
$\cal{M}$$_b$ and $\cal{M}$$_t$. Note that ``$|P'\rangle$ is
obtained from a calculation and not a physical measurement,
because Bob's particle has already been measured in the first
step of the verification phase''. But, as pointed out above, from
$\cal{M}$$_a$, $\cal{M}$$_b$, and $\cal{M}$$_t$, Bob might obtain
a $|P'\rangle$ different from $|P\rangle$ even for valid
messages. Note that the result of a measurement ($\cal{M}$$_t$)
on a quantum state ($\sigma_Z|P\rangle$) does not completely
identify the original state. In fact, one may even assume that the
arbitrator does not measure his particle of the GHZ state.
Instead, he sends it to Bob in place of the parameter
$\cal{M}$$_t$. Unfortunately, we end up again with the problem of
comparing two unknown quantum states.
This comparison test
can produce the acceptance of forged messages with non negligible
probability.

So far we have shown that the quantum signature scheme proposed 
in Ref.~\cite{ZENG_2002} is unable to guarantee
security against a dishonest Bob or a possible attacker in the 
natural scenario where $|P\rangle$ is only known to the signer Alice. 
Moreover,
we have shown that, even in the absence of dishonest parties,
this scheme, as originally proposed, does not allow Bob to recover
the message $|P\rangle$ sent by Alice. 
After having access to this
manuscript, Zeng and Keitel acknowledged that in their work they
need to ``{\it reasonably} assume that Alice, Bob, and the
arbitrator know the message $|P\rangle$'' beforehand
\cite{ZENG_PRIVATE}. Unfortunately, this crucial point for their
scheme is not mentioned at all in their original manuscript, 
and it constitutes a severe limitation
for the possible applicability of this protocol in a practical
communication scenario. 
With this strong assumption, now one
could modify the protocol in Ref.~\cite{ZENG_2002} and substitute
the parameter $\cal{M}$$_t$ by the original GHZ particle of the
arbitrator such that Bob can obtain $|P'\rangle$ (from his
knowledge of $\cal{M}$$_a$ and $\cal{M}$$_b$) and compare it with
the known $|P\rangle$. Moreover, in the signing phase Alice would
not need to send Bob the quantum state $|P\rangle$ anymore, but
only its signature $|S\rangle$. However, it seems to us that this
scheme would be rather inefficient and expensive in terms of the
quantum resources needed to perform this particular task. In the
literature there are already unconditionally secure 
classical signature schemes to sign
classical information, with or without arbitrator, that moreover
consider the natural scenario where the message to be signed does
not need to be publicly known beforehand \cite{CHAUM_1991}. 
In fact, if we assume
the availability of a trusted arbitrator, Alice and Bob could as
well use classical message authentication codes \cite{WEGMAN} to
sign their messages \cite{MACS_sign}. Therefore, we believe that
in this case it would be necessary that the authors of
Ref.~\cite{ZENG_2002} clarify
the relevance of their scheme in a practical communication
scenario together with its real advantages, if any, with respect
to unconditionally secure classical signature protocols.

The authors wish to thank P. Raynal and G. O. Myhr for very
useful discussions and comments, 
and one anonymous referee for his helpful suggestions. 
We would also like to thank G.
Zeng and C.~H. Keitel for the clarifying discussion of their
investigations. M.C. specially thanks H.-K. Lo for hospitality
and support during his stay at the University of Toronto where
this manuscript was finished. This work was supported by the DFG
under the Emmy Noether programme, the European Commission
(Integrated Project SECOQC).


\bibliographystyle{apsrev}
\bibliographystyle{apsrev}

\begin{thebibliography}{99}
\bibitem{note1} Note that, in this scenario, if one assumes 
moreover that 
Alice and Bob have preagreed on a given set of possible quantum 
states to be sent, Alice could always employ 
an unconditionally secure classical signature scheme to effectively 
sign a quantum state even if Bob does not know its identity. For that, Alice 
could sign, for instance, a  
classical label identifying the state within the preagreed set of them 
and then she sends this 
label to Bob. With this information Bob could reconstruct 
the state.
\end{thebibliography}

\end{document}